\documentstyle[prl,aps,multicol,epsf]{revtex}

\begin{document}
\draft

\title{Localization and absence of Breit-Wigner form for Cauchy 
random band matrices}

\author{Klaus M. Frahm}

\address {Laboratoire de Physique Quantique, UMR 5626 du CNRS, 
Universit\'e Paul Sabatier, F-31062 Toulouse Cedex 4, France}

\date{submitted May 24, 2000}

\maketitle

\begin{abstract}
We analytically calculate the local density of states for Cauchy random 
band matrices with strongly fluctuating diagonal elements. 
The Breit-Wigner form for ordinary band matrices is replaced by a 
Levy distribution of index $\mu=1/2$ and the characteristic energy scale 
$\alpha$ is strongly enhanced as compared to the Breit-Wigner width. 
The unperturbed eigenstates decay according to the non-exponential 
law $\propto e^{-\sqrt{\alpha t}}$. 
We analytically determine the localization length by 
a new method to derive the supersymmetric non-linear $\sigma$ model 
for this type of band matrices. 
\end{abstract}


\vspace{-0.3cm}

\begin{multicols}{2}
\narrowtext
Band matrices with random elements appear in a varity of physical problems 
in the context of quantum chaos and localization 
\cite{kicked_rotor,iwz,fyodorov0,efetov_book,guhr1}. 
A detailed analytical investigation of the properties of such matrices
was performed by Fyodorov and Mirlin \cite{fyodorov0} using 
Efetov's supersymmetry technique \cite{efetov_book}. Later, 
motivated by the localization problem of two interacting particles, 
Shepelyansky introduced and studied random band matrices superimposed 
with strongly fluctuating diagonal matrix elements \cite{shep1}. Subsequent 
work on this type of matrices \cite{band_ord} showed the existence 
of the {\em Breit-Wigner regime} where the eigenstates 
have a very peaked structure inside an eventual localization domaine. 

In all these cases the matrix elements were drawn from a regular, typically 
gaussian, distribution with finite variance. In the present work, we 
extend the band matrix ensembles of refs. \cite{shep1,band_ord} to the 
case of Cauchy distributed matrix elements. In an early work, Lloyd 
\cite{lloyd} already introduced and studied a model with diagonal Cauchy 
distributed disorder. 
A detailed study of the level statistics of full random matrices \cite{cizeau}
with elements drawn from the more general Levy distribution 
\cite{levy_dist} was performed by Cizeau and Bouchaud. Band matrix 
ensembles with Cauchy distributed matrix elements were recently argued 
to be relevant and studied numerically in the context of the localization 
problem of two interacting particles \cite{oppen1,dima_moriond}. 
In this work, we present analytical results concerning the wave function 
properties for a model similar as in ref. \cite{dima_moriond} 
but which differs from that of ref. \cite{cizeau} by two important 
features: it concerns a banded and not full matrix, and its diagonal elements 
are typically much larger than the off-diagonal coupling elements. 

Actually, we think that this case is of particular interest 
for a generic disordered fermionic many particle problem 
(for instance recently studied in \cite{many_fermions}) where 
non-interacting eigenstates are coupled by quasi-random two-particle 
interaction matrix elements. There is an interesting and subtle 
regime \cite{breit_wig_fermions} where the effective 
two-particle level spacing is smaller than the typical interaction 
matrix element while typical eigenstates are nevertheless composed of 
many non-interacting eigenstates. 
In this regime, we can diagonalize the problem for two 
particles by perturbation theory and apply the corresponding unitary 
transformation to the full many-particle Hamiltonian. This provides a new 
type of random matrix with residual three-body interaction matrix elements 
typically given by perturbative expressions with denominators, containing 
independent random non-interacting energy differences. Iterating this 
procedure one also obtains higher order contributions. Flambaum et al. 
\cite{flambaum} have argued that such denominators give effective 
distributions for these matrix elements characterized by long 
tails with the same power law as the Cauchy distribution. Even though 
the real situation is more complicated, this argument indeed shows the 
relevance of Cauchy random matrix ensembles. 

The model we consider is a random band matrix of dimension $N\times N$ and 
of width $b\gg 1$ whose matrix elements 
are of the form $H_{kl}=\eta_k\,\delta_{kl}+U_{kl}$ 
with $U_{kl}=0$ if $k=l$ or $|k-l|>b$. 
These matrix elements are statistically independent and distributed 
according to a Cauchy distribution $p_{a}(x)\equiv\pi^{-1} a/(a^2+x^2)$ 
where the width $a$ is given in terms of two 
parameters $W$ and $U_0$ via $a\equiv W/\pi$ for $\eta_k$ and 
$a\equiv U_0$ for $U_{kl}$. In this paper, we 
consider the orthogonal symmetry class where $H$ is real symmetric. The 
generalization to the other symmetry classes is straightforward. For the 
following analytical calculations, we will furthermore concentrate on 
the case where $U_0$ is 
sufficiently large to avoid a simple perturbative situation but still 
so small that the total density of states will essentially be 
dominated by the unperturbed diagonal elements only. 

For ordinary band matrices 
\cite{band_ord} with a finite variance of $U_{jk}$, this case corresponds 
to the Breit-Wigner regime where the eigenstates are in addition to 
an eventual space localization also localized in the (unperturbed) energy 
space, i.~e. only 
the sites $j$ with $|E-\eta_j|\lesssim \Gamma$ essentially contribute 
to an eigenstate of energy $E$ where $\Gamma\approx 2\pi (2b) \langle 
|U_{jk}|^2\rangle/W\ll W$ is the Breit-Wigner width. This can be seen 
in the local density of states at site $j$ which is a Lorentzian in 
$(E-\eta_j)$ of width $\Gamma/2$ \cite{band_ord}. 

To study the Cauchy band matrix ensemble introduced above, we 
therefore start with this quantity, 
\begin{equation}
\label{eq3}
\rho_j(E)=-\frac{1}{\pi}\, \mbox{Im} \left\langle 
G_{jj}^{(+)}(E)\right\rangle_j\ ,
\end{equation}
where $G^{(\pm)}(E)=(E\pm i0-H)^{-1}$ and the average 
$\langle \cdots\rangle_j$ is taken with respect to all random matrix 
elements except the diagonal element $\eta_j$ at the site $j$ under 
consideration. As in the original Lloyd model \cite{lloyd} the average 
over the other diagonal elements $\eta_k$, $k\neq j$ can be exactly performed 
by replacing in the definition of the Green function these elements 
according to: $\eta_k+i\,0\to i\,W/\pi$. 
Using a general algebraic identity for the block inverse of a matrix, 
we obtain
\begin{eqnarray}
\label{eq4}
\rho_j(E)&=&-\frac{1}{\pi}\,  \mbox{Im} \left\langle 
\frac{1}{E+i0-\eta_j+i\Gamma_j(U)/2}\right\rangle_U\ ,\\
\label{eq5}
\Gamma_j(U)&=& 2i\sum_{k,l(\neq j)}U_{jk}\, \tilde G_{kl}^{(+)}\,U_{lj}\ ,
\end{eqnarray}
where $\langle \cdots\rangle_U$ denotes the average over $U_{jk}$ only, and 
$\tilde G^{(+)}=(E+ i\,W/\pi-\tilde U)^{-1}$ with $\tilde U$ being the 
$(N-1)\times (N-1)$-matrix 
obtained from $U$ after elimination of the $j$-th row and $j$-th column. 
Eq. (\ref{eq4}) reproduces the Lorentzian Breit-Wigner form of 
width Re$[\Gamma_j(U)]/2$ for a given realization of the coupling matrix $U$. 
For ordinary band matrices this width is a self-averaging 
quantity and the further $U$-average does not modify the Breit-Wigner form. 
However, for the case of Cauchy band matrices $\Gamma_j(U)$ is strongly 
fluctuating and the $U$-average will considerably modify the 
Breit-Wigner form. Now, we consider 
the limit $|E|\ll W$ and we neglect $\tilde U$ in the definition of 
$\tilde G^{(+)}$ which is possible for the regime we consider. Using, 
$\tilde G^{(+)}\approx (-i\,\pi/W)\openone_N$, we obtain 
\begin{equation}
\label{eq6}
\rho_j(E)=\frac{1}{\pi}\,\mbox{Im}\left\{\, i\,
\int_0^\infty dt\ e^{i\,t(E-\eta_j)}\ \left[\phi\left(\frac{2\pi\,U_0^2\,t}{W}
\right)\right]^{2b}\right\}\ ,
\end{equation}
with the function $\phi(z)$ defined by
\begin{equation}
\label{eq7}
\phi(z)\equiv\frac{1}{\pi}\int_{-\infty}^\infty du\,\frac{1}{1+u^2}
\ e^{-z u^2/2}\ ,\ z\ge 0\ .
\end{equation}
In the limit $b\gg 1$, the integral (\ref{eq6}) is dominated by the behavior 
at small $t$. Using the approximation
\hbox{$\ln \phi(z)\approx -\sqrt{\frac{2}{\pi}\,z}+{\cal O}(z)$}, we obtain 
\begin{eqnarray}
\label{eq8}
\rho_j(E)&=&
\frac{1}{\alpha}L_{1/2}\left(\frac{E-\eta_j}{\alpha}\right)
\quad,\quad \alpha=\frac{16\,U_0^2\,b^2}{W}\ ,\\
\label{eq9}
L_{\mu}(s)&=&\frac{1}{2\pi}\int_{-\infty}^\infty dt\ e^{its-|t|^\mu}\ .
\end{eqnarray}
Here $L_\mu(s)$ represents a Levy distribution of index $\mu$ \cite{levy_dist} 
with the behavior $L_\mu(s)\propto s^{-1-\mu}$ for $|s|\gg 1$. 
The expressions (\ref{eq8}) and (\ref{eq9}) provide the first important 
results of this work. They show that the local density of states 
is still a peaked function of $(E-\eta_j)$ but with two important 
modifications. First, the Lorentzian Breit-Wigner form is replaced by 
the Levy distribution $L_{1/2}$, and second, the characteristic energy scale 
behaves as $\alpha\approx (4/\pi)\, b\, \Gamma\gg \Gamma$ where 
$\Gamma$ is the Breit-Wigner 
width for ordinary band matrices (with a finite variance 
$\langle U_{jk}^2\rangle\equiv U_0^2$). 
We mention that the approximation to neglect the contribution of 
$\tilde U$ in $\tilde G^{(+)}$ is valid for $\alpha\ll W$, i.~e. 
$U_0\,b\ll W$.

The first modification has an 
important implication for the time-evolution of a state $|\psi(t)\!>$ 
initially localized at one site, $|\psi(0)\!>=|j\!>$. 
Then, the average of the amplitude $<\!\psi(t)|j\!>$ obeys a 
non-exponential decay law \cite{decay_discuss} of the form 
$\propto e^{-\sqrt{\alpha\,t}}$ instead of 
$\propto e^{-\Gamma\,t/2}$. This type of behavior was for 
instance recently found in the context of many-body effects in cold 
Rydberg atoms \cite{akulin}. We conclude that it is a very general feature 
due to the long tail distribution of residual interaction matrix 
elements in many-body problems \cite{flambaum}. 

The second modification concerning the large enhancement of the 
characteristic energy scale can be qualitatively understood by the fact 
that for a given realization of $U$ the sum in eq. (\ref{eq5}) is 
essentially determined by the typical maximal 
value $2b\,U_0$ of the $2b$ different matrix elements $U_{jk}$. 

It is pretty obvious to generalize eqs. (\ref{eq8}) and (\ref{eq9}) 
to the case where the matrix elements $U_{jk}$ are drawn from the more 
general Levy distribution $U_0^{-1}\,L_\mu(U_{jk}/U_0)$. 
In this case, we have to replace in (\ref{eq8}) $L_{1/2}$ by 
$L_{\mu/2}$ and the characteristic energy scale is enhanced according 
to: $\alpha\sim b^{2/\mu-1}\,\Gamma$. 

We now turn to a field theoretical formulation of the Cauchy band matrix 
model in terms of a supersymmetric non-linear $\sigma$ model 
\cite{efetov_book}. This formulation provides the access to 
the correlators of 
an advanced Green function $G^{(+)}$ and a retarded Green function 
$G^{(-)}$ which are for example useful to study the transport 
properties, the level statistics or the appearance of dynamical localization. 
For this, typically, the ensemble average 
of a generating functional of the form
\begin{equation}
\label{eq10}
F(J)=\mbox{Sdet}^{-1/2}\left(E+({\textstyle \frac{\omega}{2}}+i\varepsilon)
\Lambda-H\otimes \openone_8+J\right)
\end{equation}
is considered. For the details concerning the supersymmetry method and the 
notations, we refer to the standard literature \cite{efetov_book,guhr1}. 
Here, we only remind that 
$\Lambda=\openone_N\otimes\mbox{diag}(1,1,1,1,-1,-1,-1,-1)$ is the 
$8N\times 8N$ 
supermatrix whose eigenvalues distinguish between the matrix blocks associated 
to the advanced and retarded Green functions and that $J$ is a source matrix 
which is used to generate the Green function correlators by taking 
suitable derivatives with respect to its matrix elements 
\cite{efetov_book,guhr1}. 

Normally, $H$ contains a certain number of gaussian random variables and 
it is possible to perform the ensemble average analytically which provides 
after a some transformations and approximations the non-linear 
supermatrix model \cite{iwz,fyodorov0,efetov_book,guhr1,band_ord}. 
For the case of Cauchy band matrices, this approach has to be modified 
because of the non-gaussian distribution of its matrix elements. Our 
strategy will be to perform analytically by a new method the 
ensemble average with respect to the Cauchy distributed diagonal 
matrix elements $\eta_j$ while the coupling matrix $U$ is kept 
fixed. This will provide a $\sigma$ model formulation for an arbitrary 
coupling matrix $U$. The ensemble average with respect to $U$ can then 
be performed at a later stage when possible and convenient. As a first 
application of this procedure, we will derive an analytic 
expression for the localization length of the Cauchy band matrix model. 

The main idea is to replace the diagonal matrix elements 
$\eta_j$ by random phases $e^{i2\varphi_j}$ via 
$\eta_j=(W/\pi)\,\tan \varphi_j$. These phases are uniformly 
distributed on the unit cercle and can be formally identified with the 
scattering matrix of a chaotic cavity ideally coupled to one scattering 
channel. Describing the Hamiltonian of the chaotic cavity by an 
$M\times M$ random matrix $h_j$ drawn from the gaussian orthogonal ensemble 
(GOE) one can apply the following substitution \cite{lewenkopf}~:
\begin{equation}
\label{eq11}
\eta_j=(W/\pi)\,\tan \varphi_j=(W/\pi)\,A_j^\dagger\, h_j^{-1}\,A_j
\end{equation}
where $A_j$ is an $M$-dimensional vector with components 
$(A_j)_l=\delta_{1,l}$ and $h_j$ is a real symmetric gaussian random matrix 
with variance $\langle (h_j)_{lk}^2\rangle=(1+\delta_{lk})/M$. The 
substitution (\ref{eq11}) is valid in the limit $M\to\infty$ which will 
be taken at the end of the calculations. We mention that this approach 
is related to recent work concerning the derivation of the 
$\sigma$ model by performing phase averages in models involving 
unitary quantum maps for classically chaotic systems 
such as the kicked rotator \cite{zirn} or rough billiards \cite{klaus}. 
The replacement (\ref{eq11}) and the following steps follow very closely 
the method described in ref. \cite{klaus}. 

Inserting (\ref{eq11}) in the generating functional (\ref{eq10}) and 
applying suitable transformations inside the superdeterminant, one 
obtains an effective model defined on a Hilbert space of dimension 
$N\cdot M$ and characterized by an effective Hamiltonian 
containing the GOE-matrices $h_j$ in its diagonal blocks and 
$(-W/\pi)\,A_j (D^{-1})_{jk}\,A_k^\dagger$ in its off-diagonal $jk$-block. 
Here $D$ is the supermatrix appearing inside the superdeterminant of 
(\ref{eq10}) with $\eta_j$ put to zero. This model is quite similar to the 
IWZ-model used in ref. \cite{iwz} and following the standard procedure 
to derive the $\sigma$ model, we obtain 
$\langle F(J)\rangle_{\eta} = \int DQ\ e^{-{\cal L}(Q)}$ with the action 
\begin{eqnarray}
\label{eq13}
{\cal L}(Q)&=& \frac{1}{2}\mbox{Str}\ \ln
\Bigl({\textstyle E+(\frac{\omega}{2}+i\varepsilon)
\Lambda-U
+i\frac{W}{\pi}\,Q+J}\Bigr)\ ,
\end{eqnarray}
where $Q$ contains in its diagonal blocks $8\times 8$ supermatrices $Q_j$ 
belonging to Efetov's coset space for the orthogonal symmetry class 
\cite{efetov_book}. We mention  
some important points concerning this result~: (i) it is exact since the 
saddle-point approximation to derive the $\sigma$ model becomes exact in 
the limit $M\to\infty$; (ii) its range of applicability goes far beyond 
the particular model considered here since it is valid for an arbitrary 
coupling matrix $U$ and its derivation only needs the Cauchy distribution 
for the diagonal matrix elements; (iii) it generalizes the above mentioned 
replacement rule for one-point functions to the case of two-point functions~: 
$\eta_k+i0\Lambda\to i\,(W/\pi)\,Q_k$ where $Q_k$ is now a dynamical 
variable over which we have to integrate. 

To analyze the localization properties, we may put $\omega=0$
and expand the action (\ref{eq13}) for the long wave length limit. 
In the regime $U_0\,b\ll W$, this gives apart from the source term 
the action
\begin{equation}
\label{eq14}
{\cal L}(Q)\approx -\frac{1}{16}\int dx\  \xi_x(U)\,
\mbox{Str}[(\partial_x Q(x))^2]
\end{equation}
with the quantity $\xi_x(U)\equiv (4\pi^2/W^2) \sum_{j=1}^b j^2\,U_{x,x+j}^2$. 
If $\xi_x(U)$ were independent of $x$ the action (\ref{eq14}) 
would correspond to the standard $\sigma$ model for quasi one-dimensional 
wires with the localization length $\xi\equiv \xi_x(U)$ 
\cite{efetov_book,guhr1}. 
This is actually the case for gaussian distributed $U_{xy}$ for which 
$\xi_x(U)\approx \langle \xi_x(U)\rangle_U=4\pi^2 b^3 U_0^2/3W^2$ is 
a selfaveraging quantity. However, for the Cauchy band matrix model 
$\xi_x(U)$ is strongly fluctuating and its average does not even exist. It is 
therefore necessary to take the $x$-dependence properly into account. 
A similar situation has recently been encountered by Rupp et al. who 
derived the 1d~$\sigma$ model with an $x$-dependent ``diffusion constant'' 
from a hierachical random-matrix model for many-body states \cite{rupp}. 
According to this work all quantities that can be extracted from the 
1d~$\sigma$ model only depend on the rescaled length variable 
$s(x)=\int_0^x dy\,\xi_y^{-1}(U)$. This justifies the intuitive average 
procedure 
$\xi^{-1}\equiv \lim_{L\to\infty} \frac{1}{L}\int_0^L dx\ \xi_x^{-1}(U)\approx \langle \xi_x^{-1}(U)\rangle$. 
Fortunately, this average is finite 
and well defined for the Cauchy band matrix case (if $b\ge 3$). 
It can be easily evaluated by the same techniques used above 
for the local density of states [see eqs. (\ref{eq4})-(\ref{eq9})]. 
For $b\gg 1$, we obtain~:
\begin{equation}
\label{eq17}
\xi=\frac{2\pi\,b^4\,U_0^2}{W^2}=b\,\gamma\quad,\quad
\gamma\equiv\frac{\pi}{16}\left(\frac{\alpha}{\Delta}\right)
\end{equation}
where the dimensionless parameter $\gamma$ counts the number of well coupled 
levels inside a strip of typical length $b$, $\Delta= W/2b$ is 
the effective level spacing of such a strip, and $\alpha$ is 
the characteristic energy scale for the local density of states (\ref{eq8}). 
(For the more general case of Levy distributed $U_{jk}$, we find:  
$\xi\sim b^{2+2/\mu}\, U_0^2/W^2$.) The relation (\ref{eq17}) compares 
to $\xi\sim b\,(\Gamma/\Delta)$ for the case of ordinary band matrices 
\cite{band_ord,dima_moriond}. However, it contradicts 
the expression $\xi\sim b \sqrt{\gamma}$ which was with proper translation 
of notations numerically obtained from 
a similar Cauchy band matrix model in ref. \cite{dima_moriond}. 
We attribute this to the extreme 
numerical difficulty to access the regime $1\ll \gamma\ll b$ 
where we expect (\ref{eq17}) to be valid. For $\gamma\lesssim 1$ we 
enter the perturbative regime with 
$\xi\sim b/\ln(\gamma^{-1})$ while for $\gamma\gtrsim b$ the localization 
length is likely to saturate at $\xi\sim b^2$. 
\vspace{-0.2cm}
\begin{figure}
\epsfxsize=0.7\hsize
\epsffile{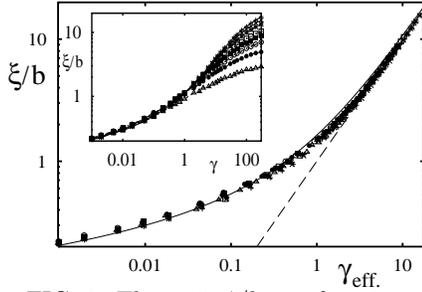}
\caption{The ratio $\xi/b$ as a function of the effective parameter 
$\gamma_{\rm eff.}$ defined in eq. (\ref{eq18}). The data points are 
numerical values, 
the dashed line corresponds to the limit $\xi/b\approx \gamma_{\rm eff.}$ 
for $\gamma_{\rm eff.}\gg 1$ and the full line is the scaling curve 
(\ref{eq18}). The inset shows $\xi/b$ versus $\gamma$. 
The different curves correspond to the values $b=10,20,30,40,50,70,85,100$ 
with $b=10$ for the lowest curve. }
\label{fig1}
\end{figure}
\vspace{-0.3cm}
To verify this picture, we have therefore numerically determined the 
localization length by the recursive Green function method for 
$0.001\le \gamma \le 300$ and $10\le b\le 100$. The inset of Fig. 
\ref{fig1} shows the ratio $\xi/b$ as a function of $\gamma$. At first sight, 
it is indeed very difficult to verify the analytical result (\ref{eq17}) and 
a direct numerical fit (e.~g. for $b=100$ and $2\le\gamma\le 30$) gives the 
power law $\xi/b\approx \gamma^{0.6}$ rather than $\xi/b\approx \gamma$. 
To understand this, we note that the number of well mixed levels inside 
one bandwidth obviously saturates at $b$ for $\gamma>b$. It appears 
therefore reasonable to replace in (\ref{eq17}) $\gamma$ by
an effective value $\gamma_{\rm eff.}$ that interpolates between 
$\gamma$ and $b$. Actually, we find that our numerical data can be 
well approximated by the scaling curve (see Fig. \ref{fig1}) 
\begin{equation}
\label{eq18}
\frac{\xi}{b}\approx \frac{1.5}{\ln(1+1.5/\gamma_{\rm eff.})}\ ,
\ \gamma_{\rm eff.}\equiv\frac{\gamma}{[1+\sqrt{3.2\,(\gamma/b)}\,]^2}
\end{equation}
which reproduces the perturbative limit for $\gamma\ll 1$, the 
analytical result (\ref{eq17}) for $1\ll \gamma\ll b$ and the 
behavior $\xi\sim b^2$ for $\gamma\gg b$. Note that eq. (\ref{eq18}) 
provides for the case $\gamma/b\ll 1$ relative large corrections 
of order $\sqrt{\gamma/b}$ explaining the numerical difficulties 
to clearly identify this regime. 

In summary, we have obtained analytical results for the local density 
of states and the localization length for Cauchy random band matrices. 
We find that the Breit-Wigner form is replaced by a Levy distribution 
with index $\mu={1/2}$ and that the Breit-Wigner width $\Gamma$ 
for ordinary band matrices is replaced by a new enhanced energy scale 
$\alpha\sim b\,\Gamma$. This energy scale also determines the localization 
length $\xi\sim b\,(\alpha/\Delta)$. From the technical point of view, 
we have derived a $\sigma$ model formulation for arbitrary matrix 
ensembles with Cauchy distributed diagonal disorder. 

The author thanks G. Caldara, B. Georgeot, J. Lages and D. Shepelyansky 
for useful discussions.

\end{multicols}


\begin{thebibliography}{99}
\bibitem{decay_discuss} We emphasize that the average amplitude (or 
average local density of states) does not capture any quantum coherence 
effects such as dynamical localization. These are seen for instance 
at the average of the absolute square of the amplitude and only at very 
long time scales comparable to the Heisenberg time.
\end{thebibliography}
\end{document}